\documentclass{article}
\usepackage{graphicx}
\usepackage{epsfig}
\def \BE {\begin{equation}}
\def \EE {\end{equation}}
\def \BEAH {\begin{eqnarray*}}
\def \EEAH {\end{eqnarray*}}
\def \BEA {\begin{eqnarray}}
\def \EEA {\end{eqnarray}}
\def \BDM {\begin{displaymath}}
\def \EDM {\end{displaymath}}

\def \eqv4d {\stackrel{4D}{\Longleftrightarrow }}

\def \bl {\mbox{\boldmath{$\ell$}}}

\def \bn {\mbox{\boldmath{$n$}}}

\def \bm #1 {\mbox{\boldmath{$m^{(#1)}$}}}
\def \hbm #1 {\mbox{\boldmath{$\hat m^{(#1)}$}}}

\def\vell {\bf{\ell}}
\def\vm {\bf{m}}
\def\vn {\bf{n}}
\def\bo {\cal{B}}
\def \bm {{\bf m}}

\def \tT{\bf{T}}

\begin{document}

\vspace*{3cm}\begin{center}{\bf \Large Classification of Higher Dimensional Spacetimes}
\\\vspace*{2cm}

{\bf A. Coley}\footnote{E-mail:  {\tt aac@mathstat.dal.ca}}
%{\bf
%M. Heller}\footnote{E-mail:  {\tt mheller@mathstat.dal.ca}}
and {\bf N. Pelavas} \footnote{E-mail:  {\tt
pelavas@mathstat.dal.ca}}\vspace*{1cm}\\{\it Department of
Mathematics and Statistics, Dalhousie University, Halifax,
NS, Canada B3H 3J5}\\[1cm]\end{center}

\begin{abstract}

We algebraically classify some higher dimensional spacetimes,
including  a number of vacuum solutions of the Einstein field
equations which can represent higher dimensional black holes. We
discuss some consequences of this work.

%We intend to keep an active version of this
%paper, together with periodic updates, on the arXiv.

\end{abstract}

\newpage

\section{Introduction}

Many higher dimensional spacetimes are now known, including a
number of vacuum solutions of the Einstein field equations which
can represent higher dimensional black holes. These N-dimensional
(ND) black holes are of physical interest, particularly in view of
the development of string theory. It is of importance to classify
these higher dimensional spacetimes algebraically
\cite{DVSI,classif}.

Higher dimensional generalizations of the Schwarzschild solution,
the Schwarzschild-Tangherlini (ST) solution \cite{tang}, which are
spherically symmetry on spacelike $(N-2)$-surfaces, are of
algebraic Weyl type D \cite{DVSI,classif}. Higher dimensional
generalizations of Reissner-Nordstrom black holes are also of type
D \cite{DeSmet2}.

A class of 5D Kaluza-Klein vacuum solutions \cite{soliton} are
also of physical interest. As we shall see, the non-black hole
solutions (i.e., all solutions except the 5D generalized
Schwarzschild solution) are not of type D (but of a more general
algebraic Weyl type). A related class of non-static spherically
symmetric solutions \cite{AbolC} is also of type D.

The Myers-Perry solution in five and higher dimensions
\cite{Myers}, a direct generalization of the 4D aymptotically
flat, rotating black hole Kerr solution, is also of type D
\cite{classif}. A class of higher dimensional Kerr-(anti) de
Sitter solutions, which are given in N-dimensions and have
$(N-1)/2$ independent rotation parameters, have been given in
Kerr-Schild form \cite{page}. These rotating black hole solutions
with a non-zero cosmological constant reduce to the 5D solution of
\cite{hht} and the Kerr-de Sitter spacetime in 4D, and the
Myers-Perry solution in the absence of a cosmological constant.

Non-rotating uncharged black string  Randall-Sundrum braneworlds
were first discussed in \cite{chamblim}. The rotating black ring
solutions (``black rings" -- BR) are vacuum, asymptotically flat,
stationary black hole solutions of topology $S^ 1 \times S^2$
\cite{ER-PRL}. These solution have subsequently been generalized
to the non-supersymmetric black ring solutions of minimal
supergravity in \cite{EEF}. There are also supersymmetric rotating
black holes that exist in five dimensions. There is the extremal
charged rotating BMPV black hole of \cite{BMPV} in minimal
supergravity, with a horizon of spherical topology (see also
\cite{DeSmet3}). The first supersymmetric black ring (solution of
5D minimal supergravity) was presented in \cite{Elvangetal04} (and
subsequently generalized in
\cite{Elvangetal05,BenWar04,GauGut05,Getal}).

There are many other higher dimensional spacetimes of interest. A
class of dimensional relativistic gyratons (RG) \cite{FIZ}, which
are vacuum solutions of the Einstein equations of the generalized
Kundt class (representing a beam pulse of spinning radiation), are
of type III.

\subsubsection{Black Hole Uniqueness}

Many of these higher-dimensional black hole solutions are of
particular physical interest, especially regarding black hole
uniqueness. In the static case, the unique asymptotically flat
vacuum black hole is the N-dimensional Schwarzschild-Tangherlini
solution \cite{GIS} (when the assumption of asymptotic flatness is
dropped, uniqueness fails even with same topology as in ST).

 In more than four
dimensions, even for pure gravity, stationary five-dimensional
black holes are not uniquely characterized by their asymptotic
conserved charges, such as mass and angular momenta. In
particular, the vacuum solutions of asymptotically flat rotating
black rings with event horizon homeomorphic to $S^ 1 \times S^2$
\cite{ER-PRL} have the same conserved charges as the stationary
Myers-Perry rotating black hole; if electromagnetic fields are
included, then the black rings can carry charge and, moreover, the
number of parameters required to specify black ring solutions now
exceeds the number of conserved quantities that they carry
\cite{HORO}. However, the $U(1)\times U(1)$  supersymmetric
solutions are only specified by a finite number of parameters.
Recently, Bena and Warner \cite{BenWar04} have studied a family of
supersymmetric solutions of five-dimensional supergravity that is
specified by seven arbitrary functions of one variable. However,
it has been argued that the only solutions which are smooth belong
to the original $U(1)\times U(1)$ invariant family (or
superpositions thereof \cite{GauGut05}), and for this reason the
Bena and Warner \cite{BenWar04} solution cannot be extended
through a $C^2$ horizon \cite{HorowitzJE}.

We begin with a brief review of the classification procedure. We
then present the classification of several important classes of
spacetimes (more details will be presented in \cite{matt}). These
results are summarized in the table in the last section. We
conclude with a brief discussion. A bibliography is appended.

\newpage

\section{Classification of the Weyl Tensor}

The algebraic classification of the Weyl tensor in higher
dimensional Lorentzian manifolds is achieved by characterizing
algebraically special Weyl tensors by means of the existence of
aligned null vectors of various orders of alignment
\cite{classif}. We consider a null frame
$\vell=\vm_0,\;\vn=\vm_1,\;\vm_2,...\vm_i$ ($\vell,\;\vn$ null
with  $\ell^a \ell_a = n^a n_a = 0, \ell^a n_a = 1$, $\vm^i$ real
and spacelike with $m_i{}^a m_j{}_a = \delta_{ij}$; all other
products vanish) in an  $N$-dimensional Lorentz-signature
space(time), so that $g_{ab} = 2l_{(a}n_{b)} +  \delta_{jk} m^j_a
m^k_b$. Indices $a,b,c$ range from $0$ to $N-1$, and space-like
indices $i,j,k$ also indicate a null-frame, but vary from $2$ to
$N-1$ only. The frame is covariant relative to the group of linear
Lorentz transformations. There are {\em null rotations} about
$\vn$ and $\vell$ and {\em spins} of the spatial frame vectors
$\vm_i$. In particular, a {\em boost} is a transformation of the
form
\begin{equation}
  \label{eq:boost}
    \hat{\vn}= \lambda^{-1}\vn,\quad
    \hat{\vm}_i=  \vm_i,\quad
    \hat{\vell}=  \lambda\, \vell, \quad \lambda \neq 0.
\end{equation}

Let $T_{a_1... a_p}$ be a rank $p$ tensor.  For a fixed list of
indices $A_1,...,A_p$, we call the corresponding $T_{A_1... A_p}$
a null-frame scalar.  These scalars transform under a boost
(\ref{eq:boost}) according to
\begin{equation}
  \label{eq:boostxform}
  \hat{T}_{A_1... A_p}= \lambda^b\,
  T_{A_1... A_p},\quad b=b_{A_1}+...+b_{A_p},
\end{equation}
where $b_0=1,\quad b_i=0,\quad b_1=-1$. We call the
above $b$ the boost-weight of the scalar.
We define the {\em boost order} of the tensor $\tT$ to be
the boost weight of its leading term.
Introducing the notation \BE
T_{\{pqrs\}}\equiv\frac{1}{2}(T_{[ab][cd]}+
 T_{[cd][ab]}), \EE
 the components of the
Weyl tensor can be decomposed and sorted by boost weight \cite{classif}:

\begin{eqnarray}
 C_{abcd} = {\overbrace{4C_{0i0j} n_{\{a}{m^i}_b n_c {m^j}_{d\}}
}}^2 + {\overbrace{8C_{010i} n_{\{a} \ell_b n_c {m^i}_{d\}}
 + 4C_{0ijk} n_{\{a} {m^i}_b {m^j}_c {m^k}_d\}}}^1 + && \nonumber \\
 \left\{\begin{array}{l}
 4C_{0101} n_{\{a} \ell_b n_c \ell_{d\}} + 4C_{01ij}
 n_{\{a} \ell_b {m^i}_c {m^j}_{d\}} +\label{eqnweyl} \\
 8C_{0i1j} n_{\{a} {m^i}_b \ell_c {m^j}_{d \}} +
 C_{ijkl} m^i_{\{a} {m^j}_b  {m^k}_c {m^l}_{d\}}\end{array}
 \right\}^0 + && \\
 {\overbrace{8C_{101i} \ell_{\{a} {n}_b \ell
 _c {m^i}_{d\}} + 4C_{1_{ijk}} \ell_{\{a} {m^i}_b {m^j}_c {m^k}_{d \}}}}^{-1}
 + {\overbrace{4C_{1i1j} \ell_{\{ a} {m^i}_b \ell_c {m^j}_{d\}}}}^{-2}.&& \nonumber
 \end{eqnarray}

The Weyl  tensor is generically of boost order $2$. If all
$C_{0i0j}$ vanish, but some $C_{010i}$, or $C_{0ijk}$ do not, then
the boost order is $1$, etc. A null rotation about $\vell$ fixes
the leading terms of a tensor, while boosts and spins subject the
leading terms to an invertible transformation.  It follows that
the boost order of a tensor is a function of the null direction
$\vell$ (only). We shall therefore denote boost order by
$\bo(\vell)$ \cite{classif}. We will {\em define} a null vector
$\vell$ to be {\em aligned} with the Weyl tensor whenever
$\bo(\vell)\leq 1$ (and we shall refer to $\vell$ as a Weyl
aligned null direction (WAND)). We will call the integer
$1-\bo(\vell)\in \{0,1,2,3\}$ the order of alignment. We will say
that the {\bf principal type} of a Lorentzian manifold is {\bf I,
II, III, N} according to whether there exists an aligned $\vell$
of alignment order $0,1,2,3$ (i.e. $\bo(\vell)=1,0,-1,-2$),
respectively. If no aligned $\vell$ exists we will say that the
manifold is of (general) type {\bf G}. If the Weyl tensor
vanishes, we will say that the manifold is of type {\bf O}. The
algebraically special types are summarized as follows:
\begin{eqnarray}
&&  Type ~~{\bf I}: ~~ C_{0i0j}=0 \nonumber\\
&&  Type~~ {\bf II}: ~~C_{0i0j}=C_{0ijk}=0 \nonumber \\
&&  Type ~~{\bf III}:  ~~C_{0i0j}=C_{0ijk}=C_{ijkl} =C_{01ij}=0 \nonumber\\
&&  Type ~~{\bf N}:  ~~C_{0i0j}=C_{0ijk}=C_{ijkl} = C_{01ij}=C_{1ijk}=0
\end{eqnarray}

 Further categorization can be obtained by
specifying {\em alignment type} \cite{classif}, whereby we try to
normalize the form of the Weyl tensor by choosing both $\vell$ and
$\vn$ in order to set the maximum number of leading and trailing
null frame scalars to zero. Let $\vell$ be a WAND whose order of
alignment is as large as possible. We then define the principal (or primary)
alignment type of the tensor to be $b_{max} - b(\vell)$. Supposing
such a WAND $\vell$ exists, we then let $\vn$ be a null vector of
maximal alignment subject to $\ell_{a} n^{a}=1$. We define the
secondary alignment type of the tensor to be $b_{max}-b(\vn)$. The
alignment type of the Weyl tensor is then the pair consisting of
the principal and secondary alignment type
 \cite{classif}. In general, for types ${\bf I}, {\bf II},
{\bf III}$ there does not exist a secondary aligned $\vn$ (in
contrast to the situation in 4D), whence the
alignment type consists solely of the principal alignment type.
Alignment types (1,1), (2,1) and (3,1) therefore form
algebraically special subclasses of types ${\bf I}, {\bf II}, {\bf
III}$ respectively (denoted types ${\bf I}_i, {\bf II}_i, {\bf
III}_i$). There is one final subclass possible, namely type (2,2)
which is a further specialization of type (2,1); we shall denote
this as type ${\bf II}_{ii}$ or simply as type ${\bf D}$.
Therefore, a type ${\bf D}$ Weyl tensor in canonical form has no
terms of boost weights $2,1,-1,-2$ (i.e., all terms are of boost
weight zero for type ${\bf D}$).

In the case in which the Weyl tensor is reducible, it is possible
to obtain more information by decomposing the Weyl tensor and
classifying its irreducible parts. In a full classification it is
necessary to count aligned directions, the dimension of the
alignment variety, and the multiplicity of principle directions
\cite{classif}. In most applications \cite{soliton,RT,brane} the
Weyl classification is relatively simple and the details of the
more complete classification are not necessary. In \cite{classif}
it was shown that the present classification reduces to the
classical 4D Petrov classification.

\subsection{Necessary conditions for WANDs}

It would be useful to be able to find a more practical way of
determining the Weyl type, such as for example employing certain
higher dimensional scalar invariants. A set of necessary
conditions for various classes, which can significantly simplify
the search for WANDs, can be given \cite{PP}:

\BEA
%\begin{center}
\begin{tabular}{rcl}
$\ell^b \ell^c \ell_{[e}C_{a]bc[d}\ell_{f]}=0$ & $\Longleftarrow  $ & $\ell$ is WAND, at most primary type I;\\
$\ell^b \ell^c C_{abc[d}\ell_{e]}=0$& $\Longleftarrow  $& $\ell$ is WAND, at most primary  type II;\\
$ \ell^c C_{abc[d}\ell_{e]}=0$ & $\Longleftarrow  $ & $\ell$ is WAND, at most primary type III;\\
$\ell^c C_{abcd}=0$ & $\Longleftarrow  $ & $\ell$ is WAND, at most primary type N.\\
\end{tabular}
\label{HDnecessity}
%\end{center}
\EEA
For type I, equivalence holds in arbitrary dimension. However, this is not the case
for more special types.

\subsection{Methods}

Therefore, there are essentially three methods currently available
to determine the Weyl type. In a straightforward approach the
alignment equations are studied, which are $\frac{1}{2}N(N-3)$
degree-4 polynomial equations in $(N-2)$ variables (and are
generally overdetermined and hence have no solutions for $N>4$), to
determine if there exist non-trivial solutions. A second method,
in which the necessary conditions are investigated, is more
practical (and results in studying essentially the same equations
but in a more organized form). This approach is followed in
classifying the Black Ring solutions (see below \cite{PP}).
Finally, in many applications which are simple generalizations of
4D solutions in which the preferred 4D null frame is explicitly
known, the 5D null frame can be guessed directly. We will begin
with some examples of this latter approach.

More details of the calculations will be presented in \cite{matt}.
It is clear that the current method of finding WANDs is very
cumbersome. It is consequently important to derive a more
practical method for determining Weyl type; for example, by
utilizing invariants as in the case of 4D \cite{kramer}.

\newpage

\section{Weyl types of some 5D vacuum spacetimes}
A number of solutions of the five dimensional vacuum Einstein
field equations are known, some of which represent higher
dimensional black hole solutions.  A null rotation about $n$ and a
null rotation about $\ell$, subgroups of the Lorentz group, yield
primary and secondary classifications, since positive and negative
boost weight components can only be made to vanish using null
rotations about $n$/$\ell$.
Explicitly, a boost is given by

\begin{equation}
\begin{array}{lll}
\hat{\ell}=b_{1}\ell, & \hat{n}=b_{1}^{-1}n, & \hat{m}_{i}=m_{i}.
\end{array}
\end{equation}
A null rotation about $n$ is given by

\begin{equation}
\begin{array}{lll}
\hat{\ell}=\ell-\frac{1}{2}\delta^{ij}d_{i}d_{j}n+d_{i}m^{i}, & \hat{n}=n, & \hat{m}_{i}=m_{i}-d_{i}n.
\end{array}
\end{equation}
A null rotation about $l$ is given by

\begin{equation}
\begin{array}{lll}
\hat{\ell}=\ell, & \hat{n}=n-\frac{1}{2}\delta^{ij}c_{i}c_{j}\ell+c_{i}m^{i}, & \hat{m}_{i}=m_{i}-c_{i}\ell.
\end{array}
\end{equation}

In the following examples the null coframe has the form

\begin{equation}
\begin{array}{ccccc}
\ell=A^2dt+ABdr, & n=\frac{1}{2}(-dt+\frac{B}{A}dr), &
m_{1}=Crd\theta, &
m_{2}=Cr\sin\theta d\phi, & m_{3}=Ddy  \label{coframe} \\

\end{array}
\end{equation}
with the functions $A,B,C$ and $D$ being specified in each case.
The corresponding metric is
given by

\begin{equation}
ds^2=-A^2dt^2+ B^2{dr^2}+C^2(r^{2}d\theta^2+r^{2}sin^{2}\theta\,
d\phi^2) + D^{2}dy^2.
\end{equation}

\subsection{5D Schwarzschild: ST} The 4D Schwarzschild
solution is spherically symmetric on spacelike 2-surfaces; an
obvious generalization to five dimensions is spherical symmetry on
spacelike 3-surfaces.  We let $y$ be a cyclic coordinate and
set (in (\ref{coframe}))

\begin{equation}
\begin{array}{llllll}
A=\left(1-\frac{2M}{r^2}\right)^{1/2}, & B=\left(1-\frac{2M}{r^2}\right)^{-1/2}, & C=1, & D=r\sin\theta \sin\phi.\\
\label{schw}
\end{array}
\end{equation}
This solution is the unique asymptotically flat static black hole
solution in 5D. It follows immediately, for this null frame, that
the Weyl basis components all have boost weight zero and therefore
this spacetime is of type D. Higher dimensional generalizations of
the Schwarzschild solution \cite{tang} are also of type D.

\subsection{Sorkin-Gross-Perry-Davidson-Owen soliton: GP} Another generalization of
the Schwarzschild solution is obtained by setting

\begin{equation}
\begin{array}{llllll}
A=\left(\frac{ar-1}{ar+1}\right)^{\epsilon\kappa}, &
B=\frac{(a^2r^2-1)}{a^2r^2}\left(\frac{ar+1}{ar-1}\right)^{\epsilon(\kappa-1)}, & C=B, &
D=\left(\frac{ar+1}{ar-1}\right)^{\epsilon}.\\          \label{soliton}
\end{array}
\end{equation}
To ensure that this is a vacuum soliton solution \cite{soliton} the consistency relation,
$\epsilon^2(\kappa^2-\kappa+1)=1$, must be satisfied. In the limit
as $\epsilon \rightarrow 0$, $\kappa \rightarrow \infty$ while
$\epsilon\kappa \rightarrow 1$, we obtain the special case $S^*$ in which
the hypersurface $y=const.$ gives the 4D
Schwarzschild solution in isotropic coordinates; we find that $S^*$ is of type D.

In the null frame given by (\ref{coframe}), the Weyl tensor has
components with boost weight +2,0,-2.  Performing a null rotation
about $n$ shows that with arbitrary $\epsilon$ and $a$ no solution
exists for $d_{i}$ such that the Weyl basis components with boost
weight +2 and +1 vanish.  However, we note that for
$\epsilon^{2}=1/3$ and $\kappa=-1$ (special case $GP_s$) a null
rotation about $n$ can be found that will make these positive
boost weight Weyl components vanish, namely

\begin{equation}
\begin{array}{ll}
d_{1}=d_{2}=0, & d_{3}=\pm 2 \left(\frac{ar+1}{ar-1}\right)^{\epsilon}, \\  \label{snc21d3}
\end{array}
\end{equation}
resulting in a primary classification of type II. Using
$\epsilon^{2}=1/3$ and $\kappa=-1$ we can then perform a null
rotation about $\ell$.  We then find that the boost weight -1 and -2
Weyl basis components can be made to vanish with the following
parameters

\begin{equation}
\begin{array}{ll}
c_{1}=c_{2}=0, & c_{3}=\pm \frac{1}{2} \left(\frac{ar-1}{ar+1}\right)^{\epsilon}. \\  \label{snc21c3}
\end{array}
\end{equation}
Therefore, the special case $GP_s$ of (\ref{soliton}), where
$\epsilon^{2}=1/3$ and $\kappa=-1$, is of type D.

Returning to the general case of arbitrary $\epsilon$ and $a$, we
have shown that a null rotation about $n$ cannot give type II.
However, by choosing $d_{1}=d_{2}=0$ and $d_{3}$ as a solution of

\begin{equation}
d_{3}^{4}+8\left(\frac{ar-1}{ar+1}\right)^{2\epsilon\kappa}\left(1+2\kappa-\frac{4ar\epsilon\kappa(1+\kappa)}{a^2r^2
+2ar\epsilon+1}\right)d_{3}^2+16\left(\frac{ar-1}{ar+1}\right)^{4\epsilon\kappa}=0,
\label{snc2s}
\end{equation}
we see that the GP metric is of type I.

\subsection{Non-static spherically symmetric solution: AC} The
following solution, obtained by Abolghasem and Coley \cite{AbolC}, is a
spherically symmetric vacuum solution containing two arbitrary
functions $\tilde{A}(t,y)$ and $\tilde{C}(t,y)$.  By an
appropriate specification of these functions it immediately
follows that this solution contains the type D GP
solution given above.  In the null frame of (\ref{coframe}), the
solution is

\begin{equation}
\begin{array}{llllll}
A=\left(\frac{1-\frac{m}{2r}}{1+\frac{m}{2r}}\right)^{-\frac{1}{\sqrt{3}}}\tilde{A}(t,y), &
B=\left(1+\frac{m}{2r}\right)^{2}\left(\frac{1-\frac{m}{2r}}{1+\frac{m}{2r}}\right)^{1+\frac{2}{\sqrt{3}}}, & C=B, &
D=\left(\frac{1-\frac{m}{2r}}{1+\frac{m}{2r}}\right)^{-\frac{1}{\sqrt{3}}}\tilde{C}(t,y),\\          \label{aac}
\end{array}
\end{equation}
where $\tilde{A}$ and $\tilde{C}$ satisfy
$(\tilde{C}^{-1}\tilde{A}_{y})_{y}=(\tilde{A}^{-1}\tilde{C}_{t})_{t}$
for vacuum.  The non-vanishing Weyl tensor basis components have
boost weights +2,0,-2. Interestingly, the vacuum condition implies
that the Weyl basis components contain $\tilde{A}$, but are
independent of $\tilde{C}$ or any derivatives of $\tilde{A}$ and
$\tilde{C}$.  Moreover, $\tilde{A}^2/\tilde{A}^{-2}$ appears only
in the boost weight +2/-2 components; therefore, by a boost
$b_{1}=\tilde{A}^{-1}$, we can transform away any occurrence of the
arbitrary function, $\tilde{A}$, in the Weyl tensor.  Next, we
perform a null rotation about $n$, using

\begin{equation}
\begin{array}{ll}
d_{1}=d_{2}=0, & d_{3}=2 \left(\frac{2r+m}{2r-m}\right)^{\frac{1}{\sqrt{3}}}, \\  \label{sndaac}
\end{array}
\end{equation}
to eliminate boost weight +2 and +1 Weyl components.  This is
followed by a null rotation about $\ell$, given by

\begin{equation}
\begin{array}{ll}
c_{1}=c_{2}=0, & c_{3}=\frac{1}{2} \left(\frac{2r-m}{2r+m}\right)^{\frac{1}{\sqrt{3}}}, \\  \label{sncaac}
\end{array}
\end{equation}
to eliminate boost weight -2 and -1 Weyl components.  In this new
frame we are left with a Weyl tensor containing only boost weight
0 components; therefore, AC is of Weyl type D.

\section{Higher dimensional  Kerr-(anti) de Sitter solutions: K(A)S}

A class of rotating black hole solutions with a non-zero
cosmological constant, which have $(N-1)/2$ independent rotation
parameters and reduce to the 5D solution of \cite{hht} and the
Kerr-de Sitter solution in 4D and the Myers-Perry solution in the
absence of a cosmological constant, have been given in Kerr-Schild
form \cite{page}.

\subsection{K(A)S is of type $D$}

In Kerr-Schild form the 5D Kerr-de Sitter metric \cite{page} is
$ds^2=d\overline{s}^2+\frac{2M}{\rho^2}\left(k_{\mu}dx^{\mu}\right)^2$, where
\begin{equation}
d\overline{s}^2=-\frac{(1-\lambda r^2)\Delta dt^2}{(1+\lambda a^2)(1+\lambda b^2)}+\frac{r^2\rho^2 dr^2}{(1-\lambda
r^2)(r^2+a^2)(r^2+b^2)}+\frac{\rho^2 d\theta^2}{\Delta}+\frac{r^2+a^2}{1+\lambda a^2}\sin^2\theta
d\phi^2+\frac{r^2+b^2}{1+\lambda b^2}\cos^2\theta d\psi^2  \label{dS}
\end{equation}
is the de Sitter metric and
\begin{eqnarray}
\rho^2 \equiv r^2+a^2\cos^2\theta+b^2\sin^2\theta, & \Delta \equiv 1+\lambda a^2\cos^2\theta + b^2\sin^2\theta.
\end{eqnarray}
The null vector $k_{\mu}$ is
\begin{equation}
k_{\mu}dx^{\mu}=\frac{\Delta dt}{(1+\lambda a^2)(1+\lambda b^2)}+\frac{r^2\rho^2 dr}{(1-\lambda r^2)(r^2+a^2)(r^2+b^2)}
-\frac{a\sin^2\theta d\phi}{1+\lambda a^2}-\frac{b\cos^2\theta d\psi}{1+\lambda b^2}.
\end{equation}
We construct the following null coframe,

\begin{equation}
\begin{array}{ccccc}
\ell=k, & n=Adt+Bdr+Jd\phi+Kd\psi, & m_{1}=\displaystyle{\frac{\rho}{\sqrt{\Delta}}d\theta}, & m_{2}=Hdt+Fd\phi, &
m_{3}=Wdt+Zd\phi+Xd\psi  \label{kds5Dnullframe}
\end{array}
\end{equation}
where
\begin{equation}
\begin{array}{cccc}
A=\displaystyle{\frac{\Delta(2Mr^2-R)}{2r^2\rho^2(1+\lambda a^2)(1+\lambda b^2)}}, &
B=\displaystyle{\frac{1}{2}+\frac{Mr^2}{R}}, & J=\displaystyle{-\frac{a\sin^2\theta(2Mr^2-R)}{2r^2\rho^2(1+\lambda
a^2)}}, & K=\displaystyle{-\frac{b\cos^2\theta(2Mr^2-R)}{2r^2\rho^2(1+\lambda b^2)}}, \nonumber
\end{array}
\end{equation}

\begin{eqnarray*}
H=-\frac{\sqrt{\Delta}(1-\lambda r^2)a\sin\theta}{(1+\lambda a^2)\sqrt{S}}, & F=\displaystyle{
\frac{\sqrt{\Delta}(r^2+a^2)\sin\theta}{(1+\lambda a^2)\sqrt{S}}}, & W=-\frac{\Delta(r^2+a^2)(1-\lambda
r^2)b\cos\theta}{r\rho(1+\lambda a^2)(1+\lambda b^2)\sqrt{S}},
\end{eqnarray*}

\begin{eqnarray*}
Z=\frac{(r^2+a^2)(1-\lambda r^2)ab\sin^2\theta\cos\theta}{r\rho(1+\lambda a^2)\sqrt{S}}, & &
X=\frac{(r^2+b^2)\cos\theta\sqrt{S}}{r\rho(1+\lambda b^2)}
\end{eqnarray*}

\noindent and we have set $R=(r^2+a^2)(r^2+b^2)(1-\lambda r^2)$, $S=\rho^2-(1-\lambda r^2)b^2\sin^2\theta$.  It turns
out that with respect to this frame the Weyl tensor has only boost weight 0 components and hence is
of type D.  In this
case the null frame in (\ref{kds5Dnullframe}) is already aligned thus eliminating the need to consider Lorentz
transformations. The method used to obtain (\ref{kds5Dnullframe}) was to first determine a null frame associated with
the de Sitter metric (\ref{dS}), under the requirement that $\ell=k$ is one of the null directions.  It then follows
that in Kerr-Schild form the Kerr-de Sitter metric will partly contain a common factor $\ell=k$, this results in a
redefinition of the null vector $n$ which will now have an $M$ dependence.

\section{Black Ring: BR}

The necessary conditions (\ref{HDnecessity}) can be used to
classify the black ring solution \cite{PP}. The Kerr solution and
the Myers-Perry solution in five dimensions are of type D with
geodesic principal null congruences. The rotating black ring
solution, which is a vacuum, asymptotically flat, stationary black
hole solution with a horizon of topology $S^1 \times S^2$, is
given in   $\{ t,\ x,\ y,\ \phi,\ \psi \} $ coordinates by
\cite{ER-PRL} \BEA
  \nonumber
  {\rm d}s^2 &=& -\frac{F(x)}{F(y)} \left( {\rm d}t+
     R\sqrt{\lambda\nu} (1 + y) {\rm d} \psi\right)^2  \\
  &&
   +\frac{R^2}{(x-y)^2}
   \left[ -F(x) \left( G(y) {\rm d}\psi^2 +
   \frac{F(y)}{G(y)} {\rm d}y^2 \right)
   + F(y)^2 \left( \frac{{\rm d}x^2}{G(x)}
   + \frac{G(x)}{F(x)} {\rm d} \phi^2\right)\right] ,
\label{ringmetric}
\EEA
where
\BE
  F(\xi) = 1 - \lambda\xi \, ,
\qquad  G(\xi) = (1 - \xi^2)(1-\nu \xi) \, .
\label{polFG}
\EE

%-----------------------------------------------
\subsection{Black ring is of type ${\rm I}_i$}

The black ring solution and its various special cases can be classified.
The method is to solve the necessity conditions and then check that these solutions do indeed
represent WANDs by calculating the components of the Weyl tensor in an appropriate frame \cite{PP}.
In order to solve the first equation in (\ref{HDnecessity}),
$\ell^a$ is denoted by  $(\alpha, \beta, \gamma, \delta, \epsilon)$.
A set of fourth order polynomial equations
in $\alpha \dots \epsilon$ is then obtained. An additional
second order equation follows from $\ell_a \ell^a=0$.
From an anlysis of these equations, it can be shown \cite{PP} that the black ring solution is algebraically special
and of type ${\rm I}_i$.

%---------------------------------------------------
\subsubsection{Black ring  is of type II on the horizon}

A transformation
leads to a metric regular on the horizon $y=1/\nu$.
The second equation in (\ref{HDnecessity}) admits a solution
$L$. It can be checked that the  boost order of the Weyl tensor in the
frame with $\bl=L$ is 0   and thus the black ring is of type II
on the horizon.

%------------------------------------------------------
\subsubsection{Myers-Perry metric is of type D}

By setting $\lambda=1$ in (\ref{ringmetric}) we obtain the Myers-Perry
metric \cite{Myers} with a single rotation parameter.
It turns out that the second equation in (\ref{HDnecessity}) admits two independent solutions
$L_{\pm}$.
When we choose a frame with $\bl \sim L_{+}$ and $\bn \sim L_{-}$ all components of the Weyl tensor with boost weights 2,1,-1,-2 vanish
and the spacetime is thus of type D \cite{DeSmet2}.

%____________________________________________

\subsection{Special Cases}
Two other special cases were considered in \cite{DeSmet}.

\subsubsection{Wrapped black string} In terms of the metric (\ref{coframe}), this
5D vacuum solution is defined by

\begin{equation}
\begin{array}{llllll}
A=\left(1-\frac{2M}{r}\right)^{1/2}, & B=\left(1-\frac{2M}{r}\right)^{-1/2}, & C=1, & D=R,\\  \label{bs}
\end{array}
\end{equation}
representing a so-called black string, wrapped around a circle of
radius $R$.  It immediately follows from the chosen coframe of
(\ref{coframe}) that the non-vanishing Weyl tensor components have
boost weight 0.  Therefore, this spacetime is of type D.  We mention in passing
that $C_{abcd}C^{abcd}=48M^2/r^6$, indicating the presence of a
singularity at $r=0$.

\subsubsection{Homogeneous wrapped object} Setting

\begin{equation}
\begin{array}{llllll}
A=\left(1-\frac{2M}{r}\right)^{1/2}, & B=1, & C=A, & D=R\left(1-\frac{2M}{r}\right)^{-1/2},\\  \label{wo}
\end{array}
\end{equation}
in (\ref{coframe}), with $R$ a constant, gives the so-called homogeneous wrapped
object  \cite{ER-PRL,DeSmet}.  This 5D vacuum solution contains singular points at $r=0$
and $r=2M$ as indicated by

\begin{equation}
C_{abcd}C^{abcd}=\frac{24M^2(2r^2-4Mr+3M^2)}{r^4(r-2M)^4}.
\end{equation}
Initially the Weyl tensor has components with boost weight
+2,0,-2.  There does not exist a null rotation about $n$ (for $r
\ne M$) that will eliminate components with boost weight +2 and
+1, but transformations exist that make only boost weight +2
components vanish, namely

%\begin{equation}
%\begin{array}{lll}
%d_{1}=0, & d_{2}=0, & d_{3}=\pm \frac{2\left[(r-2M)\left(4M-3r\pm 2\sqrt{2(r-M)(r-2M)}\right)\right]^{1/2}}{r},\\
%d_{1}=0, & d_{2}=\pm \frac{2\sqrt{2(r-M)(r-2M)}}{r} & d_{3}=\pm \frac{2i(r-2M)}{r}, \\ \label{snpel}
%d_{1}=\frac{\sqrt{8(r-M)(r-2M)-d_{2}^{2}r^2}}{r}, & d_{2}=d_{2}, & d_{3}=\pm \frac{2i(r-2M)}{r} \\
%\end{array}
%\end{equation}

\begin{eqnarray}
d_{1}=0,\hspace{0.2in}  d_{2}=0,\hspace{0.2in} d_{3}=\pm \frac{2\left[(r-2M)\left(4M-3r\pm
2\sqrt{2(r-M)(r-2M)}\right)\right]^{1/2}}{r}, \label{hwod3} \\
d_{1}=0,\hspace{0.2in} d_{2}=\pm\frac{2\sqrt{2(r-M)(r-2M)}}{r},\hspace{0.2in} d_{3}=\pm \frac{2i(r-2M)}{r},
\label{hwod21} \\
d_{1}=\pm \frac{\sqrt{8(r-M)(r-2M)-d_{2}^{2}r^2}}{r},\hspace{0.2in} d_{2}=d_{2},\hspace{0.2in} d_{3}=\pm
\frac{2i(r-2M)}{r} \label{hwod2}
\end{eqnarray}

Note that equations (\ref{hwod21}) are a special case of the
1-parameter family of solutions in (\ref{hwod2}).  Following
(\ref{hwod3}), we perform a null rotation about $\ell$ and find
that only boost weight -2 components can be made to vanish;
therefore, we obtain Weyl type I (since boost weight +2 and -2
components are now zero).  If instead we follow (\ref{hwod21}) by
a null rotation about $\ell$, we now find that both boost weight
-2 and -1 components can be made to vanish with

\begin{equation}
\begin{array}{lll}
c_{1}=\pm \frac{i}{2\sqrt{2(r-M)(r-2M)}} ,& c_{2}=\pm \frac{r}{2\sqrt{2(r-M)(r-2M)}}, & c_{3}=0. \\  \label{smc21}
\end{array}
\end{equation}
Since the non-vanishing Weyl basis components have boost weights
+1 and 0, we obtain Weyl type II$_{i}$, which is more specialized
than that found above.  We briefly mention a second specialization
of (\ref{hwod2}) with $d_{2}=0$.  In this case the solution is
similar to (\ref{hwod21}), but with $d_{1}$ and $d_{2}$
interchanged.  A null rotation about $\ell$ can then be used to
set boost weight -2 and -1 components to zero (transformation
parameters are similar to (\ref{smc21}) but with $c_{1}$ and
$c_{2}$ interchanged).  Hence this second specialization of
(\ref{hwod2}) also gives Weyl type II$_{i}$. The solution given by
(\ref{wo}) is of type II$_{i}$.

\section{Supersymmetric Black Ring: SBR}

Next, we consider supersymmetric rotating black holes that exist
in five dimensions. There is the BMPV black hole of \cite{BMPV},
with a horizon of spherical topology. The more general
supersymmetric black ring (solution of 5D minimal supergravity)
was presented in \cite{Elvangetal04,Elvangetal05}.

\subsection{BMPV is of type ${\rm I}_i$}

The BMPV metric is \cite{BMPV}

\begin{equation}
ds^2 =-f^{2}[dt-J d\phi - K d\psi]^2+f^{-2}dr^2+r^{2}(d\theta^2+\sin^2\theta d\phi^2 +
\cos^2\theta d\psi^2),
\end{equation}
where we have set

\begin{equation}
\begin{array}{lll}
f=\displaystyle{1-\frac{\mu}{r^2}} ,& J=\displaystyle{\frac{\mu\omega\sin^2\theta}{r^2-\mu}}, &
K=\displaystyle{-\frac{\mu\omega\cos^2\theta}{r^2-\mu}}. \\
\end{array}
\end{equation}

We choose the following null coframe:

\begin{equation}
\begin{array}{cc}
 \ell=f^2dt+dr-f^{2}Jd\phi-f^{2}Kd\psi, & n=\frac{1}{2}(-dt+f^{-2}dr+Jd\phi+Kd\psi),
\end{array}
\end{equation}
\begin{equation}
\begin{array}{ccc}
m_{1}=rd\theta, & m_{2}=r\sin\theta d\phi, & m_{3}=r\cos\theta d\psi.  \label{bmpvframe} \\
\end{array}
\end{equation}
The occurrence of off-diagonal metric components would suggest
that the aligned $m_{2}$ and $m_{3}$ are more complicated than
that given in (\ref{bmpvframe}).  Indeed, this frame yields Weyl
tensor components with boost weights of all orders.  We begin by
carrying out a boost with $b_{1}=(r^2-\mu)^{-1}$ to simplify the
Weyl tensor. We then perform a null rotation about $n$ and find
that the boost weight +2 components can be transformed
\footnote{At $\theta=0$ no transformation exists such  that the
boost weight +1 components can be set to zero.} to zero using

\begin{equation}
\begin{array}{ll}
d_{1}=0, & d_{3}=-d_{2}\cot\theta
\end{array}
\end{equation}
and $d_{2}$ is a root of one of the following equations
\begin{eqnarray}
\omega r^4 d_{2}^{2}-4r^3\sin\theta d_{2}+4\omega\sin^{2}\theta = 0   \label{bmpvdtr1} \\
\mu\omega r^4 d_{2}^{2}+4r^{3}(4r^2-5\mu)\sin\theta d_{2}+4\mu\omega\sin^{2}\theta =0. \label{bmpvdtr2}
\end{eqnarray}

Equations (\ref{bmpvdtr1}) and (\ref{bmpvdtr2}) each yield different canonical forms for the
Weyl tensor, thus two
cases need to be considered.  Notice that at the horizon $r=\sqrt{\mu}$ there is only one possibility since equations
(\ref{bmpvdtr1}) and (\ref{bmpvdtr2}) coincide.
Let us first consider solutions obtained from (\ref{bmpvdtr1}).  Performing a null rotation about $\ell$, we find that
boost weight -2 components can be transformed\footnotemark\ to zero using

\begin{equation}
\begin{array}{ll}
c_{1}=0, & c_{3}=-c_{2}\cot\theta            \label{bmpvc13tr}
\end{array}
\end{equation}
and $c_{2}$ is a root of the equation
\begin{eqnarray}\hspace{-1.8in}
16(r^2-\mu)(Rr-\omega)c_{2}^{3}-4r\sin\theta[2r(3\omega R-2r)(r^2-\mu)+\mu(r^2-\omega^2)]c_{2}^{2} \nonumber \\
+4r^{3}\omega\sin^{2}\theta\left[2r(r^2-\mu)-\mu(r-\omega R)\right]c_{2} - \mu r^{5}\omega^{2}\sin^{3}\theta = 0,
\hspace{-1.5in}  \label{bmpvc2tr}
\end{eqnarray}

\noindent where $R$ is determined from the root $d_{2}$ of
(\ref{bmpvdtr1}) using the relation $R=r^{2}d_{2}/(2\sin\theta)$.
This consequently shows that the BMPV metric is of Weyl type I$_i$
(\cite{DeSmet3}).

In the second case, we consider solutions obtained from (\ref{bmpvdtr2}).
Again, a null rotation about $\ell$ can be
performed to transform
\addtocounter{footnote}{-1}\footnotemark\ the boost weight -2 components to zero (the
transformation parameters $c_{1}$ and $c_{3}$ are identical to the ones given in (\ref{bmpvc13tr})).
However, $c_{2}$ is
now a root of the following equation
\begin{eqnarray}
\hspace{-1in} 8(r^2-\mu)[4Rr(r^2-\mu)-\mu(Rr-2\omega)]c_{2}^{3}+4\mu r\sin\theta[r(3\omega
R+5r)(r^2-\mu)-(r^4-\omega^{2}\mu)]c_{2}^{2} \nonumber \\
+2\mu\omega r^{3}\sin^{2}\theta[4r(r^2-\mu)-\mu(2r-\omega R)]c_{2}-\mu^{2}\omega^{2}r^{5}\sin^{3}\theta = 0,
\hspace{-0.9in}
\end{eqnarray}

\noindent and $R$ is given by $R=r^{2}d_{2}/\sin\theta$, where
$d_{2}$ is a root of (\ref{bmpvdtr2}).  In this case, as before,
the transformed frame yields Weyl type I$_i$.

\footnotetext{At $\theta=0$ no transformation exists such that the
boost weight -1 components can be set to zero.}

\subsection{SBR}
Given a class of spacetimes, such as the two parameter family of
BMPV metrics, we can define the Weyl type of the class to be the
Weyl type of the most algebraically general member contained in
the class. Since the Weyl type of any particular metric (for
example, a fixed $\mu$ and $\omega$ in BMPV) is a geometric
property determined at every point of the manifold, we generally
expect the Weyl type to vary over the manifold. Similarly, the
Weyl type of any particular metric is the Weyl type at the point
having the most algebraically general Weyl tensor.

In \cite{Elvangetal05} it was shown that the BMPV metric is a
particular case of the supersymmetric black ring (having a horizon
topology $S^{1}\times S^{2}$). It follows that the supersymmetric
black ring is at {\em most} of Weyl type I$_i$, and possibly of
type I$_G$ or $G$. The line element of the supersymmetric black
ring is \cite{Elvangetal05}
\begin{equation}
ds^2=-f^2(dt+\omega)^2+f^{-1}ds^{2}(\mathbf{R}^{4}),
\end{equation}
where
\begin{eqnarray}
f^{-1}=1+\frac{Q-q^2}{2R^2}(x-y)-\frac{q^2}{4R^2}(x^2-y^2), & \omega=\omega_{\phi}d\phi+\omega_{\psi}d\psi \\
\omega_{\phi}=-\frac{q}{8R^2}(1-x^2)[3Q-q^{2}(3+x+y)], &
\omega_{\psi}=\displaystyle{\frac{3}{2}q(1+y)+\frac{q}{8R^2}(1-y^2)[3Q-q^{2}(3+x+y)]},
\end{eqnarray}
and the four dimensional flat space is
\begin{equation}
ds^2(\mathbf{R}^{4})=\frac{R^2}{(x-y)^2}\left[\frac{dy^2}{y^2-1}+(y^2-1)d\psi^2+\frac{dx^2}{1-x^2}+(1-x^2)d\phi^2\right].
\end{equation}
Admissible coordinates values are $-1\leq x \leq 1$, $-\infty < y \leq -1$ and $\phi$,$\psi$ are $2\pi$-periodic; it is
assumed that $q>0$ and $Q \geq q^2$.

\section{Discussion}

We have  algebraically classified a number of higher dimensional
spacetimes. The results are summarized in the Table. In future, it
would be useful to classify other higher dimensional solutions,
such as other rotating black holes \cite{HORO}, higher-dimensional
C-metrics and higher-dimensional Godel spacetimes. It is also
clear that we need a more efficient way of classifying spacetimes,
perhaps in terms of scalar invariants.

The Weyl types of the BR metrics might give a hint on the Weyl
types of BR metrics that are missing (such as the doubly spinning
neutral BR) \cite{EmparanPC}. The Bena and Warner \cite{BenWar04}
family of supersymmetric solutions of 5D supergravity are
specified by seven arbitrary functions of one variable. These
solutions are specified implicitly, although an exact solution
with 3 arbitrary functions has been presented
\cite{Bena,reall,other}. We have studied various subcases of these
solutions, but we have made no substantial progress in their
classification. We had hoped that it would be possible to identify
which solutions are smooth and which are not \cite{HorowitzJE} via
their algebraic classification.

%\begin{center}
%{\bf Table}
%\end{center}

\begin{table}
\begin{center}
\begin{tabular}{|l|l|l|l|c|c|c|} \hline
&&&&&&\\
{\bf Name} & {\bf Ref} & {\bf Comments} & {\bf Type}
& {\bf ND} & {\bf Special cases} & {\bf Type}\\
&&&&&&\\ \hline
&&&&&&\\
ST & \cite{tang} & vacuum BH  & $D$ & $\surd$ &&\\
&& $R^2 \times S^3$ &&&&\\
&&&&&& \\ \hline
&&&&&&\\
GP & \cite{soliton} & vacuum soliton   & $I$ & & &\\
&& $R \times R^2 \times S^2$ &  & & & \\
&& &  & &  & \\
&& &  & & $S^*$ & $D$\\
&&&& & $GP_s$ & $D$\\
& \cite{AbolC}& && & AC & $D$.\\
&&&&&& \\ \hline
&&&&&&\\
K(A)S & \cite{page} & Rotating BH & $D$ & $\surd$ &   & \\
&& $\Lambda \neq 0$ & & &  &\\
& & & & &  & \\
& \cite{Myers}& & & & MP & $D$\\
&&&&&&\\ \hline
&&&&&&\\
BR & \cite{ER-PRL} & Rotating BR & $I_i$ &  & & \\
&& $R \times R^2 \times S^2$ & & &  &\\
&&&&&&\\
& \cite{PP} &  &  &  & $BR_H$ & $II$\\
& \cite{Myers}&  & & & MP & $D$\\
&&&&&&\\ \hline
&&&&&&\\
BMPV & \cite{BMPV} & Supersymmetric &$I_i$ &&&\\
&&&&&&\\
SBR & \cite{Elvangetal04} &&$I_i$[*]&&\\
&&&&&& \\ \hline
&&&&&&\\
VSI & \cite{DVSI} & Non BH & $N/III$ & $\surd$ &&\\
&&&&&&\\RG & \cite{FIZ} & Rel. Gyraton & $III$ & $\surd$ &&\\
&&&&&&\\
&&&&&&\\ \hline
\end{tabular}
\end{center}
\caption{The solution (name) is identified by the acronym given in the text. In the comments, features of the solution
are presented; i.e., whether it is a black hole (BH), and whether or not it is rotating, whether there is a non-zero
cosmological constant $\Lambda$, its topology etc. In the "ND" (higher dimension) column, it is indicated whether there
are higher (than 5) dimensional generalizations of the these solutions. [*] indicates that the type is at most that
specified.}
\end{table}

\newpage

Let us summarize the 4D static and stationary black hole solutions
(with topology $S^2$); there is the Schwarszchild solution, the
more general Kerr-Newman solutions, the non-vacuum
Reissner-Nordstrom spacetimes, and the non-asymptotically flat
vacuum solutions such as Schwarszchild-de Sitter spacetime. All of
these solutions have a number of symmetries, which is reflected in
their algebraic properties: namely, they are all of Weyl (Petrov)
type D \cite{kramer}. All of the known higher dimensional black
holes also have a great deal of symmetry \cite{reall}. It is
anticipated that this will again be reflected in their having
special algebraic properties. Indeed, as can be seen from the
Table, all of the  higher dimensional black holes classified here
are of algebraically special (Weyl) type. In addition, in 4D it is
known that spherically symmetric spacetimes are of type D (or O)
\cite{kramer}, and in arbitrary dimensions it has been shown that
the Weyl tensor of a spherically symmetric and static spacetime is
"boost invariant" \cite{HR} (that is, of type D). This led to a
conjecture that asserts that stationary higher dimensional black
holes, perhaps with the additional conditions of vacuum and/or
asymptotic flatness, are necessarily of Weyl type D
\cite{classif,PP}.

This conjecture has received support recently in a study of local
(so that the results may be applied to surfaces of arbitrary
topology) non-expanding null surfaces \cite{JLTP}. Assuming the
usual energy inequalities, it was found  that the vanishing of the
expansion of a null surface implies the vanishing of the shear so
that a covariant derivative is induced on each  non-expanding null
surface. The induced degenerate metric tensor, locally identified
with a  metric tensor defined on the $N-2$ dimensional  tangent
space, and the induced covariant derivative, locally characterized
by the rotation 2-form in the vacuum case, constitute the geometry
of a non-expanding null surface. The remaining components of the
surface covariant derivative lead to constraints on the induced
metric and the rotation 2-form in the vacuum extremal isolated
null surface case. This leads to the condition that at the
non-expanding horizon (i.e., the isolated null horizon) the boost
order of the null direction tangent to the surface is at most $0$,
so that the Weyl tensor is at least of type II  \cite{JLTP} (where
the {\it aligned} null vectors tangent to the surface correspond
to a double principal null direction (PND) of the Weyl tensor in
the 4D case).

\subsection{Future Work}

As in 4D, there is clearly a relationship between algebraic type
and the properties of the covariant derivative of the null
(geodesic) $\ell$ (the ${\bf L}$-tensor, defined below, which is
the higher dimensional analogue for some of the Newman-Penrose
(NP) spin coefficients). Indeed, such a relationship was exploited
in the case of type II (and D) spacetimes in the work discussed
above \cite{JLTP}, and would likely be a necessary first step in a
rigorous proof of the type-D conjecture. Ultimately, we seek a
higher-dimensional version of the Goldberg-Sachs theorem. A first
step was taken in \cite{Bianchi}, in which the Bianchi identities
in higher dimensions were studied. Here we simply  make some
comments on the properties of the ${\bf L}$-tensor for the
spacetimes that have been classified.

In a 4D vacuum space-time the Goldberg-Sachs theorem
\cite{kramer,chandra} asserts that the Weyl tensor is of type II
with repeated PND $\ell$ such that $\Psi_{0}=\Psi_{1}=0$ if and
only if (the spin coefficients) $\kappa=\sigma=0$. Given an NP
tetrad, and assuming that this result holds for both $\ell$ and
$n$, implies that the Weyl tensor is type D with repeated PND's
$\ell$ and $n$ such that $\Psi_{0}=\Psi_{1}=\Psi_{3}=\Psi_{4}=0$
if and only if $\kappa=\sigma=\nu=\lambda=0$ (the vanishing of the
spin coefficients $\nu$ and $\lambda$ indicate that $n$ is a
geodesic, shear-free congruence, respectively).
\addtocounter{footnote}{-1}\footnotemark\ For all black hole
solutions in 4D of Petrov type D, this implies that an NP tetrad
can always be chosen such that $\kappa=\sigma=\nu=\lambda=0$ and
$\Psi_{2}\neq 0$. We note that an NP tetrad for the Schwarzschild
metric can be chosen resulting in the spin coefficients
$\kappa=\sigma=\nu=\lambda=\epsilon=\pi=\tau=0$ and $\rho \neq 0$
\cite{chandra} (which immediately implies that the Schwarzschild
metric is of Petrov type D in this NP tetrad and $\Psi_{2}$ is the
only non-vanishing Weyl scalar). The condition $\epsilon=0$
implies that the null geodesic defined by $\ell$ is affinely
parametrized, and since $\rho=-\theta+i\omega$ is real, we also
have that $\ell$ is twist-free with non-vanishing expansion. In
addition, an NP tetrad for the Kerr metric giving
$\kappa=\sigma=\nu=\lambda=\epsilon=0$ can be chosen
\cite{chandra}. Therefore, $\ell$ and $n$ are null geodesic and
shear-free, and $\ell$ is affinely parametrized. Also, $\rho$ is
non zero, showing that $\ell$ is twisting with non-vanishing
expansion. Again, the Goldberg-Sachs theorem implies that Kerr is
Petrov type D in this NP tetrad and $\Psi_{2}$ is the only
non-vanishing Weyl scalar.

\footnotetext{There are generalizations of the Goldberg-Sachs
theorem to non-vacuum spacetimes \cite{kramer}.}

Let us now consider higher dimensions. Using the null frame
$e_{a}=\{\ell,n,m_{i}\}$, where $i=2,\ldots,N-1$ and assuming the
usual inner product, we have by definition

\begin{equation}
\ell_{\alpha ; \beta} \equiv L_{ab}e^{a}_{\ \alpha}e^{b}_{\
\beta}=L_{10}\ell_{\alpha} n_{\beta} + L_{11}\ell_{\alpha}
\ell_{\beta} + L_{1i}\ell_{\alpha} m^{i}_{\ \beta} +
L_{i0}m^{i}_{\ \alpha} n_{\beta} + L_{i1}m^{i}_{\ \alpha}
\ell_{\beta} + L_{ij}m^{i}_{\ \alpha} m^{j}_{\ \beta}, \label{Lij}
\end{equation}

\noindent where we have set $L_{0a}=0$ as a consequence of
$\ell_{\alpha}\ell^{\alpha}=0$ \cite{DVSI}.  Contracting
(\ref{Lij}) with $\ell^{\beta}$ gives

\begin{equation}
\ell_{\alpha;\beta}\ell^{\beta}=L_{10}\ell_{\alpha}+L_{i0}m^{i}_{\
\alpha},   \label{lgeod}
\end{equation}

\noindent from which we see that $\ell$ is geodesic if $L_{i0}=0$
and affinely parametrized if, in addition, $L_{10}=0$ ($L_{i0}$
and $L_{10}$ are the analogues of $\kappa$ and
$\epsilon+\overline{\epsilon}$, respectively). We can decompose
the purely spatial part of $L$ as $L_{ij}=S_{ij}+A_{ij}$ where
$S_{ij}=L_{(ij)}$ and $A_{ij}=L_{[ij]}$ \cite{DVSI}. Further
decomposing $S_{ij}=\sigma_{ij}+\frac{S}{N-2}\delta_{ij}$ into its
trace-free and trace parts identifies the shear and expansion of
$\ell$, respectively. From (\ref{Lij}), the expansion of $\ell$ is
given by

\begin{equation}
\theta := \frac{1}{N-2}\ell^{\alpha}_{\
;\alpha}=\frac{1}{N-2}(L_{10}+L_{ij}\delta^{ij})=\frac{1}{N-2}(L_{10}+S).
\end{equation}
Therefore, only when the null geodesic $\ell$ is also affinely
parametrized ($L_{10}=0$) can we identify $Tr(S_{ij})=S$ with the
expansion of $\ell$. \addtocounter{footnote}{-1}\footnotemark\ If
$\ell$ is geodesic we can always choose an affine parametrization
by applying an appropriate boost; consequently we shall assume
that $L_{10}=0$.

\footnotetext{We note that in the study of isolated horizons, the
space-time derivative operator induces a derivative operator on
the null hypersurface  \cite{JLTP}; it is with respect to this
derivative operator that the (projected) expansion of the null
normal $\ell$ is zero.}

Considering the 5D Schwarzschild (ST) metric and calculating
$L_{ab}$ shows that $\ell$ is geodesic but is not affinely
parametrized in the frame presented above.  Performing a boost
with $b_{1}=1/A^2$ affinely parametrizes $\ell$, resulting in
$\theta=1/r$ and $\sigma_{ij}=A_{ij}=0$.  Next we consider the
non-static spherically symmetric AC metric and find that the
transformed null vector $\ell$, the one aligned with the Weyl
tensor, is geodesic but not affinely parametrized. Performing a
boost $b_{1}=f(t,y)$, where $f$ satisfies
$(f\tilde{C})_{t}=(f\tilde{A})_{y}$, results in an affine
parametrization; it then follows from $L_{ab}$ that
$S_{ij}=A_{ij}=0$, and hence the null geodesic is expansion-free,
shear-free and twist-free.  Last, we calculate $L_{ab}$ for the 5D
Kerr-de Sitter (KS) metric in the frame considered. We find that
$\ell$ is geodesic and affinely parametrized.  The expansion of
$\ell$ is
\begin{equation}
\theta=\frac{3r^2+a^2\cos^2\theta+b^2\sin^2\theta}{3r(r^2+a^2\cos^2\theta+b^2\sin^2\theta)}.
\end{equation}
However, unlike the 4D Kerr metric, in higher dimensions we find
that $\ell$ has non-zero shear; the  shear invariants

\begin{equation}
\begin{array}{cc}
\sigma_{i}^{\ j}\sigma_{j}^{\
i}=\frac{2}{3}\frac{(a^2\cos^2\theta+b^2\sin^2\theta)^2}{r^2(r^2+a^2\cos^2\theta+b^2\sin^2\theta)^2},
& \sigma_{i}^{\ j}\sigma_{j}^{\ k}\sigma_{k}^{\
i}=\frac{2}{9}\frac{(a^2\cos^2\theta+b^2\sin^2\theta)^3}{r^3(r^2+a^2\cos^2\theta+b^2\sin^2\theta)^3},
\label{shkerr}
\end{array}
\end{equation}
are non-vanishing (unless $a=b=0$), even though this space-time is
of Weyl type D (also see \cite{Bianchi}). This implies that any
higher dimensional version of the Goldberg-Sachs theorem will
necessarily be more complicated.

\subsection{Bibliography and Updates}

 We intend to keep an active version of this
article, with periodic updates, on the arXiv. {\footnote {Authors
wishing to send us references for exact higher dimensional
solutions for us to add to the bibliography or exact solutions
that have been classified for us to add to the table(s), please
send e-mail to aac@mathstat.dal.ca with the subject header
WClass.}} In particular, in the arXiv version we will include a
bibliography of known exact higher-dimensional solutions (and
especially black hole solutions)

\newpage

\section{Acknowledgements}

This work was supported, in part, by NSERC. We would like to thank
I.~Bena, R. Milson and V. Pravda for helpful correspondence.

%\section*{References}


\begin{thebibliography}{99}



\bibitem{DVSI} A. Coley, R. Milson, N. Pelavas, V. Pravda, A. Pravdova and R. Zalaletdinov,
Phys. Rev. D.  {\bf 67}, 104020 (2003); A. Coley, R. Milson, V. Pravda and A. Pravdov\'a,  Class. Quantum Grav. {\bf
21}, 5519 (2004).

\bibitem{classif} A. Coley, R. Milson, V. Pravda and A. Pravdov\'a,
Class. Quantum Grav. {\bf 21}, L35 (2004) [gr-qc/0401008]; R.
Milson, A. Coley, V. Pravda and A. Pravdov\'a, Int. J. Geom. Meth.
Mod. Phys.  (2004) [gr-qc/0401010].

\bibitem{tang} F. R. Tangherlini, Nuovo Cim. {\bf 27}, 636 (1967).


\bibitem{DeSmet2} P-J. De Smet, Gen. Rel. Grav. {\bf 36}  1501 (2004).


\bibitem{soliton} R.D. Sorkin, Phys. Rev. Lett. {\bf 51}, 87 (1983);
D.J. Gross and M.J. Perry,  Nucl. Phys. B {\bf 226}, 29 (1983);
A. Davidson and D.A. Owen,  Phys. Lett. B {\bf 155}, 247 (1985).

\bibitem{AbolC} G. Abolghasem and A. Coley, preprint;
G. Abolghasem, A.A. Coley and D.J. McManus,
Gen. Rel. Grav. {\bf 30}, 1569 (1998).


\bibitem{Myers}
R. C. Myers and M. J. Perry,  Ann. Phys. {\bf 172}, 304 (1986).

\bibitem{page}  G. W. Gibbons, H. Lu, D.N. Page
and C. N. Pope, J. Geom. Phys.; hep-th/0404008; see also Phys.
Rev. Letts. {\bf 93}, 171102 (2004).


\bibitem{hht} S. W. Hawking, C. J. Hunter and M. M.
Taylor-Robinson, Phys. Rev. D {\bf 59}, 064005 (1999).


\bibitem{chamblim} S. Chamblin, S. W. Hawking and H.~S.~Reall,
Phys. Rev. D {\bf 65}, 084010 (2002); see also Phys. Rev. D {\bf
61}, 0605007 (2000).


\bibitem{ER-PRL}
R.~Emparan and H.~S.~Reall,
Phys.\ Rev.\ Lett.\  {\bf 88}, 101101 (2002)
[hep-th/0110260]; H.~Elvang,  Phys.\ Rev.\ D {\bf
68} 124016  (2003) [hep-th/0305247];
H.~Elvang and R.~Emparan,  JHEP {\bf 0311}
035 (2003) [hep-th/0310008];
R.~Emparan, JHEP {\bf 0403}, 064 (2004) [hep-th/0402149].

\bibitem{EEF} H.~Elvang, R.~Emparan and P. Figueras, JHEP {\bf 0502}
031 (2005) [hep-th/0412130].


\bibitem{BMPV}
J.~C.~Breckenridge, R.~C.~Myers, A.~W.~Peet and C.~Vafa,
Phys.\ Lett.\ B {\bf 391}  93 (1997)
[hep-th/9602065].


\bibitem{DeSmet3} P-J. De Smet, Gen. Rel. Grav. {\bf 37},  237 (2005).


\bibitem{Elvangetal04}
H. Elvang,  R. Emparan, D. Mateos and H.~S. Reall,
  Phys.\ Rev.\ D {\bf 71}, 024033 (2005)
  [hep-th/0408120].

\bibitem{Elvangetal05}
H. Elvang,  R. Emparan, D. Mateos and H.~S. Reall,
  Phys.\ Rev.\ Lett.\  {\bf 93}, 211302 (2004)  [hep-th/0407065].

\bibitem{BenWar04}
I. Bena and N.~P. Warner, preprint UCLA/04/TEP/31
[hep-th/0408106].


\bibitem{GauGut05}
  J.~P.~Gauntlett and J.~B.~Gutowski,
  Phys.\ Rev.\ D {\bf 71}, 045002 (2005)
  [hep-th/0408122];
  J.~P.~Gauntlett and J.~B.~Gutowski,
  Phys.\ Rev.\ D {\bf 71}, 025013 (2005)
  [hep-th/0408010].


\bibitem{Getal}
J.~P.~Gauntlett, J.~B.~Gutowski, C.~M.~Hull, S.~Pakis and
H.~S.~Reall, Class.\ Quant.\ Grav.\ {\bf 20}, 4587 (2003)
[hep-th/0209114].


\bibitem{FIZ} V. Frolov, W. Israel and A. Zelnikov,
hep-th/0506001; see also Phys. Rev. D {\bf 71}, 104034 (2005)

\bibitem{GIS} G. W. Gibbons, D. Ida and T. Shiromizu,
Phys. Rev. Letts. {\bf 89}, 041101 (2002) [gr-qc/0203004].


\bibitem{HORO} G. Horowitz,
to appear in {\em Kerr Spacetime: Rotating Black Holes"}, eds. S.
Scott, M. Visser, and D. Wiltshire (Cambridge University Press)
[gr-qc/0507080].

\bibitem{HorowitzJE}
  G.~T.~Horowitz and H.~S.~Reall,
  Class.\ Quant.\ Grav.\  {\bf 22}, 1289 (2005)
  [hep-th/0411268].


\bibitem{matt} M. Heller, Master's Thesis, Dalhousie University,
2005.


\bibitem{brane} V. Rubakov and M. Shaposhnikov,  Phys. Lett. B {\bf
125}, 139 (1983);  N. Arkani-Hamed, S. Dimopoulos  and  G. Dvali,
Phys. Lett. B {\bf 429},  263 (1998); L. Randall and  R. Sundrum,
Phys. Rev. Lett. {\bf 83}, 3370 \& 4690 (1999).

\bibitem{RT} A.~A.~Tseytlin,
Phys. Rev. D {\bf 47}, 3421 (1993); A.~A.~Tseytlin,
Nucl. Phys. B {\bf 390}, 153 (1993); G.H. Horowitz
and  A.A. Tseytlin, Phys. Rev D {\bf 51}, 2896 (1995); J.~G.~Russo and A.~A.~Tseytlin,
JHEP {\bf 0204}, 021 (2002); {\em ibid.} {\bf 0209}, 035 (2002);
M. Blau, J. Figueroa-O'Farrill, C. Hull and G. Papadopoulos
JHEP {\bf 0201},  047 (2002); {\em ibid.}
Class. Quant. Grav. {\bf19}, L87 (2002);
P. Meessen, Phys. Rev.  D {\bf 65}, 087501 (2002);
R.R. Metsaev, Nucl. Phys. {\bf B625}, 70 (2002);
R.~R.~Metsaev and A.~A.~Tseytlin,
Phys. Rev.  D{\bf 65}, 126004 (2002);
J.~Maldacena and L.~Maoz, JHEP
{\bf 0212}, 046 (2002).


\bibitem{PP}
V. Pravda and A. Pravdov\'a, Gen. Rel. Grav. {\bf 37},  1277
(2005) [gr-qc/0501003].



\bibitem{kramer} D. Kramer, H. Stephani, E. Herlt and M. A. H. MacCallum,
{\it Exact Solutions of Einstein's Field Equations} (Cambridge
University Press, Cambridge, England, 1980).


\bibitem{Bianchi} V. Pravda, A. Pravdov\'a, A. Coley and R. Milson,
Class. Quantum Grav. {\bf 21}, 2873 (2004) [gr-qc/0401013].

\bibitem{DeSmet} P-J. De Smet, Class. Quantum Grav. {\bf 19}, 4877 (2002).



\bibitem{Bena}
I.~Bena and P.~Kraus, Phys. Rev. D  {\bf 72} 025007
 (2005)
[hep-th/0503053]; P.~Kraus, and
N. P. Warner, [hep-th/0504142];
I.~Bena and P.~Kraus,
  Phys.\ Rev.\ D {\bf 70}, 046003 (2004);
 I.~Bena, C.~W.~Wang and N.~P.~Warner, [hep-th/0411072]; I.~Bena,
Phys. Rev. D {\bf 70} 105018 (2004)
[hep-th/0404073].



\bibitem{reall}
H.~S.~Reall,
Phys.\ Rev.\ D {\bf 68}, 024024 (2003) [hep-th/0211290].



\bibitem{other}
J.~B.~Gutowski and H.~S.~Reall,
  JHEP {\bf 0404}, 048 (2004) [hep-th/0401129]; R.~Emparan, D.~Mateos and P.~K.~Townsend,
JHEP {\bf 0107}, 011 (2001)
[hep-th/0106012]; M.~Cvetic and F.~Larsen,  Nucl.\
Phys.\ B {\bf 531}, 239 (1998) [hep-th/9805097];
J.~B.~Gutowski, D.~Martelli and H.~S.~Reall, Class.\
Quant.\ Grav.\  {\bf 20}, 5049 (2003)
[hep-th/0306235].


\bibitem{EmparanPC} R.~Emparan, private communication.


\bibitem{HR} G. Horowitz and S. Ross, Phys. Rev. D {\bf 57}, 1098
(1998).


\bibitem{JLTP} J. Lewandowski and T. Pawlowski,
Class.\ Quant.\ Grav.\  {\bf 22}, 1573 (2005)
[gr-qc/0410146].


\bibitem{chandra} S. Chandrasekhar, {\it The Mathematical Theory of Black Holes} (Oxford University Press, 1992).


\end{thebibliography}
\end{document}